\documentstyle[titlepage,12pt]{article}

\newread\epsffilein    
\newif\ifepsffileok    
\newif\ifepsfbbfound   
\newif\ifepsfverbose   
\newdimen\epsfxsize    
\newdimen\epsfysize    
\newdimen\epsftsize    
\newdimen\epsfrsize    
\newdimen\epsftmp      
\newdimen\pspoints     
\def\epsfbox#1{\global\def\epsfllx{72}\global\def\epsflly{72}%
   \global\def\epsfurx{540}\global\def\epsfury{720}%
   \def\lbracket{[}\def\testit{#1}\ifx\testit\lbracket
   \let\next=\epsfgetlitbb\else\let\next=\epsfnormal\fi\next{#1}}%
\def\epsfgetlitbb#1#2 #3 #4 #5]#6{\epsfgrab #2 #3 #4 #5 .\\%
   \epsfsetgraph{#6}}%
\def\epsfnormal#1{\epsfgetbb{#1}\epsfsetgraph{#1}}%
\def\epsfgetbb#1{%
\openin\epsffilein=#1
\ifeof\epsffilein\errmessage{I couldn't open #1, will ignore it}\else
   {\epsffileoktrue \chardef\other=12
    \def\do##1{\catcode`##1=\other}\dospecials \catcode`\ =10
    \loop
       \read\epsffilein to \epsffileline
       \ifeof\epsffilein\epsffileokfalse\else
          \expandafter\epsfaux\epsffileline:. \\%
       \fi
   \ifepsffileok\repeat
   \ifepsfbbfound\else
    \ifepsfverbose\message{No bounding box comment in #1; using defaults}\fi\fi
   }\closein\epsffilein\fi}%
\def\epsfsetgraph#1{%
   \epsfrsize=\epsfury\pspoints
   \advance\epsfrsize by-\epsflly\pspoints
   \epsftsize=\epsfurx\pspoints
   \advance\epsftsize by-\epsfllx\pspoints
   \epsfxsize\epsfsize\epsftsize\epsfrsize
   \ifnum\epsfxsize=0 \ifnum\epsfysize=0
      \epsfxsize=\epsftsize \epsfysize=\epsfrsize
     \else\epsftmp=\epsftsize \divide\epsftmp\epsfrsize
       \epsfxsize=\epsfysize \multiply\epsfxsize\epsftmp
       \multiply\epsftmp\epsfrsize \advance\epsftsize-\epsftmp
       \epsftmp=\epsfysize
       \loop \advance\epsftsize\epsftsize \divide\epsftmp 2
       \ifnum\epsftmp>0
          \ifnum\epsftsize<\epsfrsize\else
             \advance\epsftsize-\epsfrsize \advance\epsfxsize\epsftmp \fi
       \repeat
     \fi
   \else\epsftmp=\epsfrsize \divide\epsftmp\epsftsize
     \epsfysize=\epsfxsize \multiply\epsfysize\epsftmp
     \multiply\epsftmp\epsftsize \advance\epsfrsize-\epsftmp
     \epsftmp=\epsfxsize
     \loop \advance\epsfrsize\epsfrsize \divide\epsftmp 2
     \ifnum\epsftmp>0
        \ifnum\epsfrsize<\epsftsize\else
           \advance\epsfrsize-\epsftsize \advance\epsfysize\epsftmp \fi
     \repeat
   \fi
   \ifepsfverbose\message{#1: width=\the\epsfxsize, height=\the\epsfysize}\fi
   \epsftmp=10\epsfxsize \divide\epsftmp\pspoints
   \vbox to\epsfysize{\vfil\hbox to\epsfxsize{%
      \includegraphics{#1}%
      \hfil}}%
\epsfxsize=0pt\epsfysize=0pt}%

{\catcode`\%=12 \global\let\epsfpercent=
\long\def\epsfaux#1#2:#3\\{\ifx#1\epsfpercent
   \def\testit{#2}\ifx\testit\epsfbblit
      \epsfgrab #3 . . . \\%
      \epsffileokfalse
      \global\epsfbbfoundtrue
   \fi\else\ifx#1\par\else\epsffileokfalse\fi\fi}%
\def\epsfgrab #1 #2 #3 #4 #5\\{%
   \global\def\epsfllx{#1}\ifx\epsfllx\empty
      \epsfgrab #2 #3 #4 #5 .\\\else
   \global\def\epsflly{#2}%
   \global\def\epsfurx{#3}\global\def\epsfury{#4}\fi}%
\def\epsfsize#1#2{\epsfxsize}
\textheight=23cm
\textwidth=16cm
\setlength{\topmargin}{-1cm}
\setlength{\oddsidemargin}{0cm}
\addtolength{\textheight}{1cm}
\begin{document}
\begin{flushright}
PITHA 96/02\\
hep-ph/9602273\\
February 96 
\end{flushright}
\vspace{0.8cm}
\begin{center}
{\bf\LARGE CP Nonconservation  }
{\bf\LARGE in Top Quark Production by (Un) Polarized
$e^+e^-$ and $\gamma \gamma$ Collisions\footnote{Contribution to 
the Workshop on $e^+e^-$ Linear Colliders; Annecy, Assergi,
Hamburg, Febr. - Sept. 1995. Research 
supported  by BMBF contract 056AC92PE.} }

\vspace{2cm}
\centerline{ \bf W. Bernreuther$^{a}$, A. Brandenburg$^a$,
and P. Overmann$^b$}
\vspace{1cm}
\centerline{$^a$Institut f. Theoretische Physik,
RWTH Aachen, D-52056 Aachen, Germany}
\centerline{$^b$Institut f. Theoretische Physik,
Universit\"at Heidelberg, D-69120 Heidelberg, Germany}
\vspace{2cm}

{\bf Abstract:}\\
\parbox[t]{\textwidth}
{We report on an investigation of CP violation in (un)polarized
$e^+e^-\to{\bar t}t$  resulting from an extended neutral Higgs sector or from 
the minimal SUSY 
extension of the Standard Model (SM). We consider  c.m. energies from the
${\bar t}t$ threshold to the TeV range. In addition sensitivity estimates 
for CP-violating form factors of the top quark are made.
Further we discuss the prospects of probing Higgs sector CP violation
in $\gamma\gamma\to{\bar t}t$.}
\end{center}
\bibliographystyle{unsrt}
\section{Introduction and Summary}
\noindent In recent years quite a number of proposals 
have been made on how top quarks
can serve as probes of CP-violating interactions beyond the
Kobayashi-Maskawa (KM) mechanism (\cite{BNOS}-\cite{ABB}).
Model-independent analyses in terms of form factors can be found in
\cite{BNOS,KLY,BO,ArSe,AS,BM,Rind}.
Here we report on the results of an investigation  
of CP-violating interactions 
from an extended Higgs sector and from the minimal
supersymmetric extension of the SM and their
effects on $\bar{t}t$ production (and decay) at a linear collider.
First we consider $e^+e^-\to {\bar t}t$.
The new features of this study,
as compared to previous work \cite{BNOS,BO}, are:
we take into account the possibility of longitudinally polarized 
electron beams which enhance some of the effects, and we propose and study
 optimized observables with maximal sensitivity
to CP effects for "semileptonic" $t\bar t$ events. For the Higgs  model we 
find: (a) The highest sensitivities are reached somewhat above 
the $t\bar t$ threshold. (b) Longitudinal electron polarization would be
an asset; yet the sensitivity of our best observable depends only weakly 
on the electron polarization. In the above models the effects
in $e^+e^-\to t\bar{t}$ are due to non-resonant
radiative corrections and are therefore not easy to detect. 
If a light Higgs boson $\varphi$ with 
mass $m_{\varphi} <$ 200 GeV and sizable CP-violating couplings to top
quarks exists then there is a chance to see a signal at a collider
with 50 fb$^{-1}$ integrated luminosity. 
It would obviously be more promising to check for CP
violation directly in $\varphi$ decays \cite{BBra,DK}.
Effects in $t\bar{t}$ production
 due to a CP phase in gluino exchange
are too small to produce statistically significant signals.\\
In addition we have also investigated how well one could measure 
in the reactions (1), (2), (3) 
below the CP-violating form factors of the top quark 
with our optimized
observables. Results are given in section 2. Of course, the sensitivity to the
(dimensionful) form factors increases considerably with the c.m. 
energy and with
the availability of beam polarization. (See also \cite{BNOS,Rind}.)\\
High energetic photon-photon collisions \cite{Ginz}, which are discussed in 
the context of a linear collider, may provide, among other things,
an interesting possiblity to produce neutral Higgs bosons and  study their
quantum numbers. In the framework of two-Higgs doublet extensions
we have investigated CP violation in unpolarized 
$\gamma \gamma\to{\bar t}t$
which includes resonant Higgs boson  production and decay into ${\bar t}t$.
Here the effects can be much larger than in $e^+e^-$-annihilation.
If one or more Higgs bosons of intermediate mass 300 GeV 
$\vbox{\hbox{$\scriptstyle<$}\vskip-11pt\hbox{$\scriptstyle\sim$}} 
m_{\varphi}
\vbox{\hbox{$\scriptstyle<$}\vskip-11pt\hbox{$\scriptstyle\sim$}}$ 
500 GeV exist
we find that $\gamma \gamma\to\varphi\to {\bar t}t$ would be a promising 
channel to study its/their CP properties.
Sensitivity estimates  will be given in section 3.
For a detailed exposition of our studies and further references see
 \cite{BO2,ABB}.

\section{$e^+e^-\to {\bar t}t$}

In this section  we consider the production of a top quark pair via the
collision of
an unpolarized positron  beam and a longitudinally polarized electron beam:

\begin{equation}
 e^+ ({\bf e}_+)  +  e^- ({\bf e}_-, p)  \quad \to \quad
t( {\bf k}_t) +  \bar{t} ({\bf k}_{\bar{t}}).
\label{TheReaction}
\end{equation}

\noindent Here $p$ is the longitudinal polarization of the electron beam
($p=1$ refers to right handed electrons). For our purposes the most interesting
final states are those from semileptonic $t$ decay and non-leptonic
$\bar{t}$ decay and vice versa:

\begin{equation}
t  \enskip  \bar{t}  \quad \to \quad \ell^+({\bf q}_+) + \nu_\ell + b
   + \overline{ X}_{\rm had}({\bf q}_{\bar X}),
\label{TopDecay}
\end{equation}

\begin{equation}
  t  \enskip  \bar{t} \quad \to \quad  X_{\rm had}({\bf q}_X) +
\ell^-({\bf q}_-) +  \bar \nu_\ell     + \bar{b} \  \ ,
\label{AntiTopDecay}
\end{equation}

\noindent  where the 3-momenta in eqs. (1) - (3)
refer to the $e^+ e^-$ c.m. frame. \\
\noindent  CP-violating interactions can affect the $\bar{t}t$ production
and decay vertices. Quantum mechanical interference of the CP-even and -odd
parts
of the amplitudes for the above reactions then lead to the correlations
which we are
after. For CP-nonconserving neutral 
Higgs boson couplings  and for 
CP-nonconserving gluino-quark-squark couplings it has 
been shown {\cite{BO}} that 
these interactions lead to 
larger effects in ${\bar t}t$ production than in $t$ (and $\bar{t}$) decay.
Therefore we consider observables which are predominantly
sensitive to CP effects in the production amplitude which,  in these SM
extensions, arise at 1-loop through induced electric and weak dipole
form factors $d^{\gamma,Z}_t(s)$. The real parts
${\rm Re} d_t^{\gamma,Z}$  generate  a difference in the $t$ and $\bar{t}$
polarizations
orthogonal to the scattering plane of reaction (1), 
whereas non-zero absorptive parts
${\rm Im} d_t^{\gamma,Z}$ lead to a difference in the $t$ and $\bar{t}$
polarizations along the top direction of flight. 
The class of events (2), (3) is highly suited
to trace these spin-momentum correlations in the $\bar{t}t$ production 
vertex through final state momentum correlations:
From the hadronic momentum in (\ref{TopDecay}),(\ref{AntiTopDecay}) one can
reconstruct the $\bar{t}$ and $t$ momentum, 
respectively and hence the rest frames
of these quarks. 
The extremely short life time of the top quark implies that the top
polarization is essentially undisturbed by hadronization effects and
can be analyzed by its parity-violating weak decay $t \to b + W$, 
which we assume
to be the dominant decay mode.  Further,
the charged lepton from semileptonic top decay is known to be by far the best
analyzer of the top spin {\cite{CJK}. Therefore
our observables  are chosen to be functions of the directions of the
hadronic system from top decay, of the charged lepton momentum, of
the positron beam direction, and of the c.m. energy $\sqrt s$.
In order to increase the statistical sensitivity
of the observables  we shall use the lepton unit momenta 
$\hat{\bf q}^*_{\pm}$ in the
corresponding top rest frames, which are directly accessible 
in the processes (2), (3).\\
\noindent The CP ``asymmetries'' which we discuss
are differences of expectation values

\begin{equation}
{\cal A} \quad = \quad \langle{\cal O}_+ (s,\hat{\bf q}^*_+
,\hat{\bf q}_{\bar X}, \hat{\bf e}_+)\rangle -
              \langle{\cal O}_- (s,\hat{\bf q}^*_-
,\hat{\bf q}_{X}, \hat{\bf e}_+)\rangle \ \ ,
\label{GeneralForm}
\end{equation}

\noindent where the mean values refer to events
(\ref{TopDecay}), (\ref{AntiTopDecay}) respectively. 
For instance, $ {\cal O}_+=(\hat{\bf q}_+^*
\times\hat{\bf q}_{\bar{X}})\cdot\hat{\bf e}_+$. The observable ${\cal
O}_-$ is
defined to be the CP image of ${\cal O}_+$. It is obtained from
${\cal O}_+$
by the substitutions $\hat{\bf q}_{\bar X} \to -\hat{\bf q}_{X}$,
$\hat{\bf q}^*_+ \to -\hat{\bf q}^*_-$, $\hat{\bf e}_+ \to  \hat{\bf e}_+$.
The  ratio

\begin{equation}
 r \quad = \quad {\langle{\cal O}_+\rangle - \langle{\cal O}_- \rangle \over
 \Delta{\cal O}_+}
\label{Sensitivity}
\end{equation}

\noindent is a measure of the statistical sensitivity
of ${\cal O}$. Here $\Delta{\cal O} = \sqrt{\langle{\cal O}^2\rangle
-\langle{\cal O}\rangle^2}$. For the observables used
in this paper we have  $\Delta{\cal O} \approx \sqrt{\langle{\cal O}^2\rangle}$
and $\Delta{\cal O}_+\simeq \Delta{\cal O}_-$.  The signal-to-noise ratio
of ${\cal A}$
is given by $S_{\cal A } = |r| \sqrt N_{event} / \sqrt 2$, where 
$N_{event}$ is the number
of events
of type (2) or (3).\\
\noindent In \cite{BO2} explicit expressions are given for two observables 
${\cal O}_{\pm}(1),{\cal O}_{\pm}(2)$ with
optimized  sensitivity $|r|$, and  these observables pick up dispersive
(${\rm Re} d_t^{\gamma,Z}$) and  absorptive (${\rm Im} d_t^{\gamma,Z}$)
CP effects, respectively.  In addition to the
functional dependence exhibited in (\ref{GeneralForm}) they depend also on
the electron beam polarization $p$. As the expressions are somewhat 
lengthy we do not reproduce them here.
Schematically an optimized observable is constructed as follows:
Consider a differential cross section which is  of the form ${\rm d}
\sigma = {\rm d}\sigma_0 + \lambda {\rm d} \sigma_1$ where  $\lambda$ is a 
small parameter. One can show {\cite{AS, Diehl}} that 
the  observable with the highest statistical sensitivity 
is given by ${\cal O} = {{\rm d} \sigma_1 /{\rm d} \sigma_0 }$.\\
\noindent We shall consider phase space cuts which
are CP-symmetric.
When the $e^+e^-$ beams are unpolarized (or transversely
polarized) the
asymmetries (\ref{GeneralForm}) can be classified as being odd under a CP
transformation.
This means that contributions to $\langle {\cal O}_\pm \rangle$ from
CP--invariant interactions  cancel in the difference.
If the electron beam is longitudinally polarized the initial $e^+e^-$ state
is no longer
CP-symmetric in its c.m. frame and the CP classification no longer applies.
Contributions
from CP-conserving interactions can, in principle, contaminate ${\cal A}$
if $p\neq 0$.
However, in practice, this is not a problem because it can be argued that
SM interactions induce contaminations at the per mill level \cite{BO2}. 
On the other hand only ratios $|r|>$0.01 have a chance to be detectable 
even at a high luminosity linear collider.\par\noindent
We now  recall the salient features of neutral Higgs sector CP violation.
For definiteness we consider  two-doublet extensions of the SM with
explicit CP violation in the Yukawa couplings (which leads to the
Kobayashi-Maskawa phase) and in the Higgs potential. As a consequence
the three physical neutral Higgs boson states $\varphi_{1,2,3}$ are 
in general states 
with indefinite CP parity; i.e. they couple both to scalar and
pseudoscalar quark and lepton currents with strength
$ a_{jf} m_f/v$ and $\tilde{a}_{jf} m_f/v$, respectively, where $m_f$ is
the fermion mass and $v\simeq 246$ GeV. For the top quark we have 
$a_{jt}=d_{2j}/\sin \beta$, $\tilde{a}_{jt}=-d_{3j}\cot \beta$,
where  $\tan \beta=v_2/v_1$ is the ratio of the moduli of the 
vacuum expectation values of the two 
doublets, and $d_{2j},\ d_{3j}$ are the matrix elements 
of a $3\times 3$ orthogonal matrix which describes the mixing 
of the neutral  Higgs states of definite CP parity. 
Only the CP=+1 components of the mass eigenstates $\varphi_j$ couple to
the $W,\ Z,$ and charged Higgs bosons at the Born level. 
(For notation and details, see \cite{BSP}). 
\par 
CP violation requires that the neutral Higgs bosons are not mass-degenerate.
Only if one of the bosons, say $\varphi_1$, is rather light then there is 
a chance that effects are detectable in $e^+e^-\to{\bar t}t$. 
For the evaluation of the ratios $r$ shown in Figs.1 
we have put $m_{\varphi_1}$=100 GeV and have assumed that $m_{\varphi_{2,3}}
>> m_{\varphi_1}$. Then the effect of $\varphi_{2,3}$ on the dipole form 
factors is negligible. 
The EDM form factor and, to a good approximation \cite{BSP}, the WDM 
form factor of the top are proportional to 
$\gamma_{CP}=-a_{1t}\tilde{a}_{1t}$ which is a measure of the strength of 
CP violation induced by $\varphi_1$ exchange. One may assume maximal 
CP violation in the neutral Higgs sector by putting 
$d_{i1} = 1/\sqrt 3 (i=1,2,3)$. The product $\gamma_{CP}$ 
increases with decreasing $\tan\beta$. Phenomenologically 
the experimental upper bounds on the neutron and electron EDMs give the rough 
upper bound $|\gamma_{\rm CP}|\vbox{\hbox{$\scriptstyle<$}
\vskip-11pt\hbox{$\scriptstyle\sim$}} 5.$ Nevertheless it would 
be interesting to obtain from the top system direct information on 
$\gamma_{CP}$.\par\noindent 
In Figs.1 we have plotted the ratios $r_{1,2}$ for the optimized dispersive 
and absorptive observables ${\cal O}_{\pm}(1),{\cal O}_{\pm}(2)$ 
defined in \cite{BO2}. We have put
$m_{\rm t} = 180 \hskip 3pt{\rm GeV}$,   
$m_{\varphi_1} = 100 \hskip 3pt{\rm GeV}$, and $\gamma_{\rm CP} = 1$. 
The ratios $r_{1,2}$ are directly proportional to $\gamma_{\rm CP}$.
\par\noindent
\par
\vskip-3.5cm
\setlength{\unitlength}{1cm}
\hskip-0.7cm
\begin{minipage}{6cm}
\begin{picture}(6,10)
\epsfysize=6cm
\epsfbox{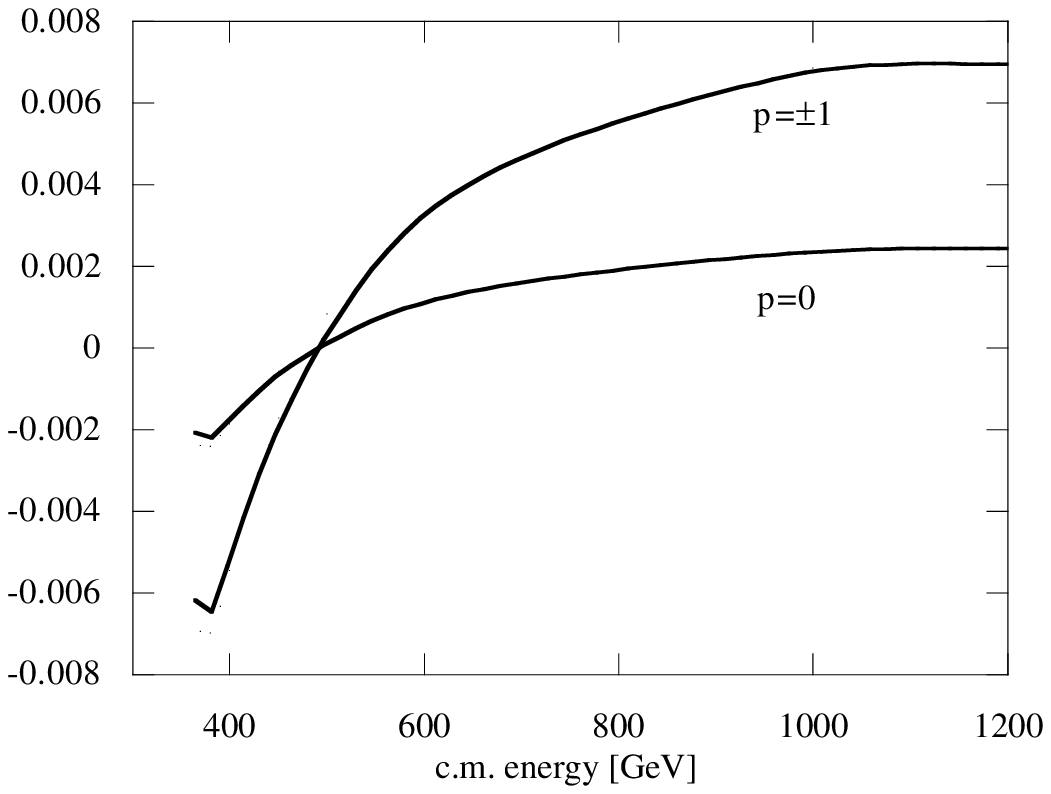}
\end{picture}
\end{minipage}\hskip 2.2cm
\begin{minipage}{6cm}
\begin{picture}(6,10)
\epsfysize=6cm
\epsfbox{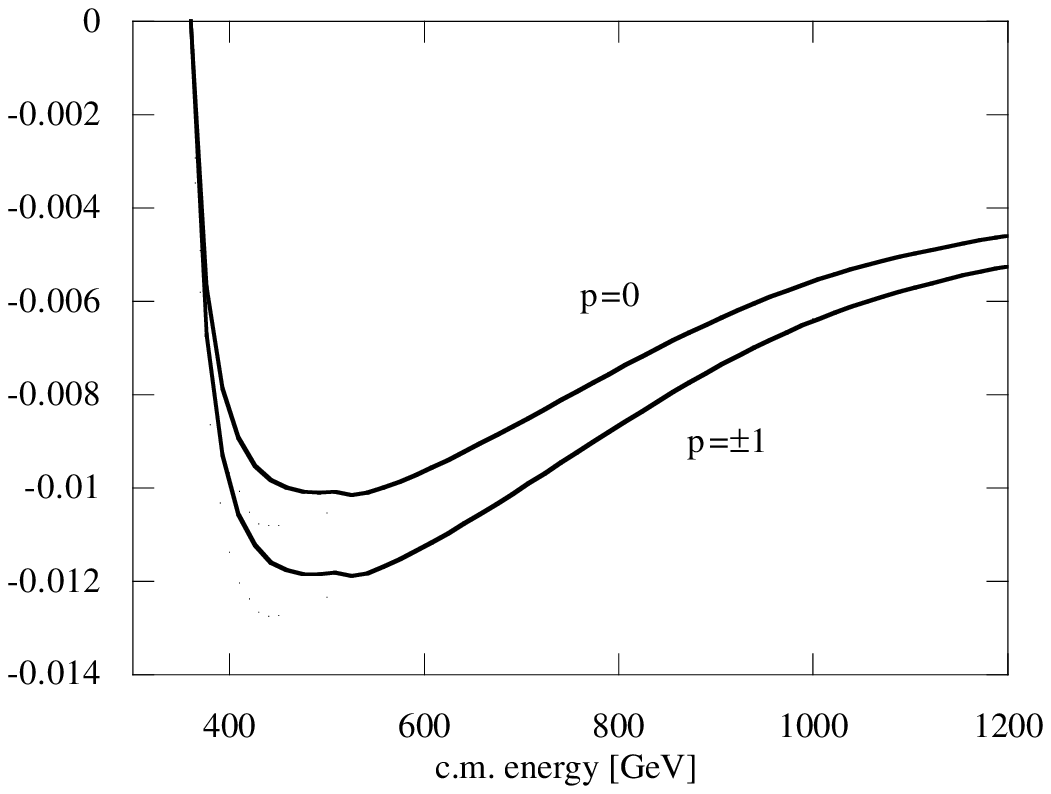}
\end{picture}
\end{minipage}
\vskip 0.3cm
{\noindent
Fig 1: \quad Ratios $r_{1}$ (left figure) and $r_{2}$ (right figure)
  for the optimized dispersive
and absorptive observables ${\cal O}_{\pm}(i)$, $i=1,2$
defined in \cite{BO2} for $m_{\rm t} =
180 \hskip 3pt{\rm GeV},$  $m_{\varphi_1} = 100 \hskip 3pt{\rm GeV},$
and $\gamma_{\rm CP} = 1$. }

\vspace{0.5cm}

\noindent Figs.1 show that the absorptive asymmetry  has the highest
sensitivity to $\gamma_{CP}$ and that it depends only weakly on the beam
polarization.
The maximal sensitivity is reached at $\sqrt s \simeq$ 450 GeV where
$|r_2|$ = 1.2$\%$ (1$\%$) if $p$=$\pm$1 (0). In order to detect this as a 3
s.d. effect
one would need 125000 (180000) events of the type (\ref{TopDecay}) and of
(\ref{AntiTopDecay}).
These event numbers are unrealistically large. If the
CP-violating effect is larger,
say $\gamma_{\rm CP}$=4, then only 1/16 of these events would be needed for
a 3 s.d. signal.
In order to reach a sensitivity $\vbox{\hbox{$\scriptstyle>$}
\vskip-11pt\hbox{$\scriptstyle\sim$}}
$ 3 s.d. to couplings $|\gamma_{\rm CP}|\vbox{\hbox{$\scriptstyle>
$}\vskip-11pt\hbox{$\scriptstyle\sim$}}$4
an integrated luminosity of
50 ${\rm fb}^{-1}$ must be collected. \par
\noindent  In general the asymmetries become smaller with increasing Higgs
mass. If we keep $\gamma_{\rm CP}$ fixed  but change the Higgs
mass to $m_{\varphi}$=200 GeV  then the maximal value at $\sqrt s \approx$ 
450 GeV
of the ratio $r_2$ in Fig.1 is reduced by about 30$\%$.\\
In summary, for light Higgs masses $m_{\varphi}<$200 GeV and sizable
CP-violating coupling $\gamma_{\rm CP}$ there is a chance to see Higgs
sector CP violation
as an induced dipole moment effect in $\bar{t}t$ production. A light Higgs
particle $\varphi$ would also
be produced at a linear collider. A consequence of $\varphi$ not being a CP
eigenstate would be a CP violation effect in the $\varphi$ 
fermion-antifermion
amplitude at Born level which could be detected in $\varphi\to\tau^+\tau^-$
\cite{BBra}.
(For further tests of the CP properties of Higgs particles, 
see \cite{AS2,DK,Kremer}.)\par\noindent
In the minimal supersymmetric extension of the SM  additional  CP--violating
phases (besides the KM phase) can be present in the Majorana mass terms, 
e.g. of the
gluinos, and in the squark (and slepton) mass matrices. For mass eigenstates  
these phases
then appear for instance in the ${\rm t}\tilde{\rm t}-\rm gluino$  
couplings 
in the form
(flavour mixing is ignored)
\begin{equation}
 {\cal L}_{\tilde {\rm t} {\rm t} \lambda} =
 i\sqrt 2 \enskip g_{\rm\scriptscriptstyle QCD} \left\{
 e^{i\phi_{\rm t}} \enskip\tilde{\rm t} _L^*  T^a ( \bar\lambda^a {\rm t}_L)
 + e^{-i\phi_{\rm t}} \enskip\tilde {\rm t}^*_R T^a (\bar \lambda^a {\rm
t}_R) \right\} +
h.c.
\label{LSusy}
\end{equation}

\noindent with $\phi_{\rm t} = \phi_\lambda - \phi_{\tilde {\rm t}}$ 
and fields
$\tilde {\rm t}_{R,L}$ which are related to the 
fields $\tilde {\rm t}_{1,2}$  
corresponding 
to mass eigenstates by an orthogonal transformation.
The 1-loop EDM and WDM form factors $d^{\gamma,Z}_t$ generated by (\ref{LSusy})
are proportional to $g^2_{QCD}\sin(2\phi_{\rm t}).$
We have computed these form factors and the resulting asymmetries 
(\ref{GeneralForm})
for the optimized observables ${\cal O}_{\pm}(1),{\cal O}_{\pm}(2)$
for maximal CP violation, $\sin(2\phi_{\rm t})=1,$ and a range of SUSY masses
 $m_{\lambda}, m_{1,2}  \ge$ 150 GeV. We have found \cite{BO2} that the 
ratios $|r_{1,2}|
\vbox{\hbox{$\scriptstyle<$}\vskip-11pt\hbox{$\scriptstyle\sim$}}$0.01.
Hence it is unlikely that a statistically significant 
signal can be detected at a 50 ${\rm  fb}^{-1}$ linear collider even for 
maximal SUSY
CP violation. (Related studies were made in \cite{Wien}.)\\
Furthermore we also investigated how well one could measure
${\rm Re} d_t^{\gamma,Z}$ and ${\rm Im} d_t^{\gamma,Z}$ independent
of any model.
For this purpose we computed the corresponding ratios $r$
for observables which, for a given polarization $p$, have optimized 
sensitivity to one of these form factors. Using an 
integrated luminosity of 20 fb$^{-1}$ (50 fb$^{-1}$) 
at $\sqrt s$ = 500 GeV (800 GeV)
we estimate the 1 s.d. statistical errors 
given in Table 1, assuming a tagging efficiency of 1 for the channels
(2), (3).
\ 
\vskip 0.8cm
\begin{tabular}{c|c|c|c||c|c|c|}
  & \multicolumn{3}{c||}{$20 {\mbox{\ fb}}^{-1},\sqrt{s}=500{\mbox{\ GeV}}$} &
\multicolumn{3}{c|} {$50 {\mbox{\ fb}}^{-1},\sqrt{s}=800{\mbox{\ GeV}}$} \\
  & $p=0$ & $p=+1$ & $p=-1$ & $p=0$ & $p=+1$ &  $p=-1$ \\
\hline
$\delta{\mbox{Re}}d_{t}^{\gamma}$ & 4.6 & 0.86 & 0.55 & 1.7 & 0.35 & 0.23 \\
\hline
$\delta{\mbox{Re}}d_{t}^{Z}$ & 1.6 & 1.6 & 1.0  & 0.91 & 0.85 & 0.55 \\
\hline 
$\delta{\mbox{Im}}d_{t}^{\gamma}$ & 1.3 & 1.0 & 0.65 & 0.57 & 0.49 & 0.32 \\
\hline
$\delta{\mbox{Im}}d_{t}^{Z}$ & 7.3 & 2.0 & 1.3  & 4.0 & 0.89 & 0.58 \\
\hline
\end{tabular}

\ 
\par 
{\noindent
Table 1: \quad 1 s.d. sensitivities to the CP-violating form factors in units
of $10^{-18}$ e cm. }

\vspace{0.5cm}
Table 1 shows that the sensitivity to ${\rm Re} d_t^{\gamma}$ and to 
${\rm Im} d_t^{Z}$ grows subtantially with beam polarization. 
For a fixed number of $t\bar{t}$ events the sensitivity to 
the nonrenormalizable ``couplings'' $ d_t^{\gamma,Z}$ 
obviously increases with increasing c.m. energy.
However, one should keep in mind that the form 
factors vary with $\sqrt s$, and 
this variation is determined by thresholds 
which depend on the underlying dynamics
of CP violation. If -- like in the  Higgs model --
the relevant branch point where the form factors develop imaginary
parts is set by the $t\bar{t}$ threshold then,
as exemplified by Fig.1, there is no gain in going
to very high energies. Needless to say: a priori one does not know.

\section{$\gamma\gamma\to {\bar t}t$}

\noindent After the discovery of a Higgs boson $\varphi$ at some collider
 a high energetic ``Compton collider'' of high luminosity could be tuned to 
$\gamma\gamma\to\varphi$ in order to perform a detailed study of the 
$\varphi$ quantum numbers (see,e.g.,\cite{Kremer}). If the photon 
polarizations 
are adjustable then one could check, as proposed in \cite{GG},
for instance with the event asymmetry $(N(+,+)-N(-,-))/(N(+,+)+N(-,-))$ 
whether or
not $\varphi$ is a CP eigenstate. (Here $\pm$ 
refer to the photon helicities.)
For unpolarized  $\gamma\gamma\to\varphi$ the CP property of $\varphi$ 
can be infered
from the final states into which it decays. If $\varphi$ is not a CP 
eigenstate  then a
CP-violating spin-spin correlation is induced in its fermionic decays 
already at the Born level,
which can be as large as 0.5 \cite{BBra}. It could be detectable in
$\varphi\to\tau^+\tau^-$ and, for sufficiently heavy $\varphi$, in 
$\varphi\to{\bar t}t$.
However, the narrow width approximation does not apply for a Higgs boson with
mass in the vicinity or 
above the $t{\bar t}$ threshold and interference with 
the non-resonant $\gamma \gamma \to t {\bar t}$ amplitude decreases this 
spin-spin
 correlation significantly. In order to include  also the case of light 
Higgs bosons $\varphi$ below the $t {\bar t}$ threshold we have 
computed, for
two-Higgs doublet extensions of the SM, the complete set of 
CP-nonconserving contributions
to $\gamma \gamma \to t {\bar t}$ in one-loop approximation \cite{ABB}.
In the spin density matrix
for ${\bar t}t$ production by unpolarized 
photon beams this leads to CP-violating dispersive
contributions with the spin structure
$\hat{\bf k}_t\cdot ({\bf s}_+\times{\bf s}_-)$  
and to absorptive contributions which are 
of the form $\hat{\bf k}_t\cdot ({\bf s}_+-{\bf s}_-)$ 
and to similar terms where $\hat{\bf k}_t
\to\hat{\bf p}_{\gamma}$. Here ${\bf s}_+,{\bf s}_-$ 
are the spin operators of $t$ and ${\bar t}$,
respectively. The latter structures correspond to polarization asymmetries.\\
As in the case of $e^+e^-\to {\bar t}t$ these spin 
correlations and polarization asymmetries
are most efficiently traced through the "semileptonic"
 ${\bar t}t$ decays (2) and the
charge conjugated channels (3). The spin-spin correlation leads
for events (2) to the triple product correlation
${\cal O}_{+} = ({\bf q}_{ +}\times \hat{{\bf q}}^*_{W^-})\cdot 
\hat{\bf q}_{\bar X}$.
Here the asterisk denotes the ${\bar t}$ rest system. 
The CP-reflected observable
${\cal O}_{-}$ applies to the charge conjugated channels (3), 
and a non zero difference
$\langle {\cal O}_{+}\rangle - \langle{\cal O}_{-}\rangle$ 
would be an unambiguous sign 
of CP violation. A CP asymmetry with a somewhat higher 
sensitivity to Higgs boson 
induced effects is

\begin{equation}
{\cal A}_{abs} = \langle {\bf q}_{+}\cdot \hat{\bf q}_{\bar X}\rangle 
- \langle 
{\bf q}_{-}\cdot \hat{\bf q}_{X}\rangle ,
\label{abs}
\end{equation}

\noindent which is a consequence of  the single spin asymmetries. \\
For the calculation of these asymmetries we used the normalized photon
distributions $N(x)$ with energy fraction $x=E_{photon}/E_{beam}$  
as given, e.g. 
in \cite{Ginz,KMS}. The maximal energy fraction of a photon $x_{max}$
is determined by the laser energy    
$x_{max} = z/(1+z)$ with $z = 4E_{beam}E_{Laser}/ m_e^2$.
For a given beam energy the laser energy is chosen such that $z$ reaches its
maximal value. This value is determined by the threshold for 
unwanted $e^+e^-$ pair production through annihilation of a backscattered
photon with a laser photon. One gets $x_{max}\simeq$ 0.8284.\\
Our result for the ratio $r$ based on the asymmetry (\ref{abs})
is shown in Fig.2 for $e^+e^-$ c.m. energy $\sqrt s$= 500 GeV and $m_t$=180 GeV
as a function of the Higgs boson mass $m_{\varphi}$ for two sets of parameters:
$d_{i1}=1/\sqrt 3$, $\tan \beta$=0.627 (set 1, solid line) and
$d_{i1}=1/\sqrt 3$, $\tan \beta$=0.3 (set 2, dashed line). These two sets
correspond to $\gamma_{CP}$=1 and $\gamma_{CP}$=3.87, respectively.
In contrast to $e^+e^-\to {\bar t}t$ we have here a complicated dependence
of the asymmetries on the Higgs boson couplings, because the 
diagrams in which the
$\varphi$ propagator can become resonant contain also bosonic loops,
and because in the resonant $\varphi$ propagator the total width 
$\Gamma_{\varphi}$
enters. The asymmetry ${\cal A}_{abs}$ has an 
extremum when the Higgs mass is 
close to the ${\bar t}t$ threshold due to large interference effects 
and an additional extremum when the Higgs mass is close to the maximal
$\gamma\gamma$ energy.  Even for a Higgs with mass above the maximal
$\gamma\gamma$ energy there remains a significant interference effect
that might be detectable. \vfil
\ \ 
\setlength{\unitlength}{1cm}
\begin{picture}(15,10)
\hskip 2.5cm 
\epsfysize=12cm
\epsfbox{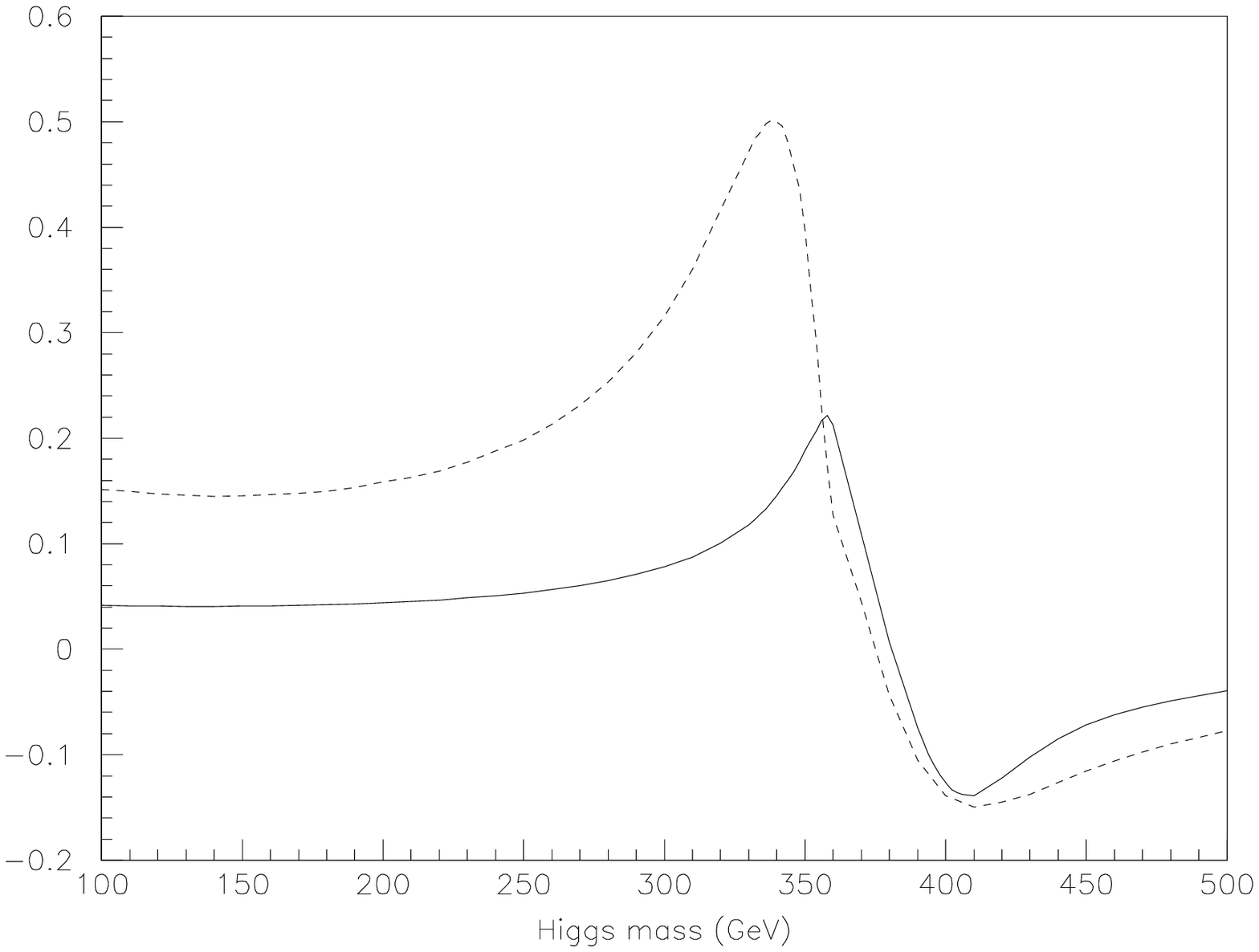}
\end{picture}
\vskip-2.8cm
{\noindent
Fig 2: \quad The ratio  $r$  for the asymmetry ${\cal A}_{abs}$ 
as a function of the Higgs mass $m_{\varphi}$ at $\sqrt s$ = 500 GeV.
The solid (dashed) line corresponds to the parameter set 1 (set 2). }

\vspace{0.5cm}

One may  also determine, for a given Higgs mass, the $e^+e^-$ collider
energy that maximizes the signal-to-noise ratio of an asymmetry.
We have done this for ${\cal A}_{abs}$ and the parameter set 1.  The results
are listed in Table 2. In brackets
we also give the numbers for set 2. (Note that $S$ is not optimized with 
respect 
to this set.) In computing the number of events of the channels (2) or the 
charge conjugated channels (3) we use as in 
sect.1 semileptonic top decays into electrons, muons, and tauons.\par
\ 

\begin{tabular}{|c|c|c|c|}\hline
$m_{\varphi}\left[{\rm GeV}\right]$ 
& $\sqrt{s_{opt}}\left[{\rm GeV}\right]$ 
& $N_{event}/\left( {\cal L}/(100 \ {\rm fb}^{-1})\right)$
& $S/\sqrt{{\cal L}/(100\ {\rm fb}^{-1})}$ 
\\ \hline\hline
 100 & 700 & $5.97(5.35)\times 10^3$ &  $1.8(5.4)$ \\ \hline
 150 & 700 & $5.96(5.31)\times 10^3$ &  $1.8(5.3)$ \\ \hline
 200 & 700 & $5.93(5.21)\times 10^3$ &  $1.9(5.5)$ \\ \hline
 250 & 680 & $5.44(4.70)\times 10^3$ &  $2.1(6.1)$ \\ \hline
 300 & 650 & $4.67(3.84)\times 10^3$ &  $2.6(7.5)$ \\ \hline
 325 & 630 & $4.10(3.30)\times 10^3$ &  $3.0(8.9)$ \\ \hline
 350 & 570 & $2.39(2.32)\times 10^3$ &  $4.0(11.4)$ \\ \hline
 375 & 600 & $3.41(4.27)\times 10^3$ &  $4.4(8.8)$ \\ \hline
 400 & 710 & $6.18(6.42)\times 10^3$ &  $3.6(7.7)$ \\ \hline
 425 & 520 & $1.87(2.31)\times 10^3$ &  $3.7(4.7)$ \\ \hline
 450 & 560 & $3.02(3.39)\times 10^3$ &  $3.4(4.5)$ \\ \hline
 475 & 580 & $3.58(3.85)\times 10^3$ &  $3.3(4.2)$ \\ \hline
 500 & 610 & $4.38(4.58)\times 10^3$ &  $2.9(3.7)$ \\ \hline
\end{tabular}

\ 
\par 
{\noindent
Table 2: \quad Optimal collider energies, number of events of 
samples (2) or (3),
and statistical significance $S$ of ${\cal A}_{abs}$ for some 
Higgs masses and parameter
set 1. The numbers in brackets are for set 2. }

\vspace{0.5cm}

\noindent Table 2 tells us, for example, that if 
$m_{\varphi}=350$ GeV (and if set 1 applies), one gets the largest 
sensitivity to the
CP asymmetry  ${\cal A}_{abs}$ at  a collider energy of $570$ GeV. 
For an integrated $e^+e^-$ luminosity of 50 fb$^{-1}$ 
(and an $e \gamma$ conversion rate of one) we then get a statistical 
significance $S$ of 2.8(8.0). By using an optimized observable instead of
(\ref{abs}) the sensitivity could still be enhanced. Should Higgs boson(s)
of intermediate mass  be discovered then, as our study shows,
$\gamma\gamma\to\varphi\to {\bar t}t$ is 
a promising channel for probing Higgs sector CP violation.

\end{document}